\documentclass{article}
\usepackage[preprint]{spconfa4}
\usepackage{siunitx,booktabs,spconfa4,amsmath,graphicx,bm,amsfonts,tikz,pgf}
\usepackage[acronym]{glossaries}
\usepackage{cite}
\usepackage{color}
\usepackage{hhline}

\usepackage{pgfplots}
\usepackage{pgfplotstable}
\usepgfplotslibrary{units}
\usetikzlibrary{pgfplots.statistics, pgfplots.colorbrewer}
\pgfplotsset{compat=1.9}
\makeatletter
\pgfplotsset{
    unit code/.code 2 args=
    \begingroup
    \protected@edef\x{\endgroup\si{#2}}\x
}
\newacronym{sma}{SMA}{spherical microphone array}
\newacronym{mlp}{MLP}{multilayer perceptron}
\newacronym{pinn}{PINN}{physics-informed neural network}
\title{A Physics-Informed Neural Network-Based Approach for the Spatial Upsampling of Spherical Microphone Arrays}
\name{\begin{minipage}{\textwidth}\centering Federico Miotello, Ferdinando Terminiello, Mirco Pezzoli,\\Alberto Bernardini, Fabio Antonacci, Augusto Sarti\end{minipage}\thanks{This work has been funded by "REPERTORIUM project. Grant agreement number 101095065. Horizon Europe. Cluster II. Culture, Creativity and Inclusive Society. Call HORIZON-CL2-2022-HERITAGE-01-02.".}
\thanks{This work was supported by the Italian Ministry of University and Research (MUR) under the National Recovery and Resilience Plan (NRRP), and by the European Union (EU) under the NextGenerationEU project.}
\thanks{This work was supported by the European Union under the Italian National Recovery and Resilience Plan (NRRP) of NextGenerationEU, partnership on ``Telecommunications of the Future'' (PE00000001---program ``RESTART'').}}
\address{Dipartimento di Elettronica, Informazione e Bioingegneria - Politecnico di Milano\\
Via Ponzio 34/5, 20133 Milano, Italia}

\toappear{\begin{minipage}{\textwidth}\footnotesize© 20XX IEEE. Personal use of this material is permitted. Permission from IEEE must be obtained for all other uses, in any current or future media, including reprinting/republishing this material for advertising or promotional purposes, creating new collective works, for resale or redistribution to servers or lists, or reuse of any copyrighted component of this work in other works.\end{minipage}}

\begin{document}
%
%
%
\def\BibTeX{{\rm B\kern-.05em{\sc i\kern-.025em b}\kern-.08em
    T\kern-.1667em\lower.7ex\hbox{E}\kern-.125emX}}
\ninept
\maketitle
\begin{abstract}
Spherical microphone arrays are convenient tools for capturing the spatial characteristics of a sound field.
However, achieving superior spatial resolution requires arrays with numerous capsules, consequently leading to expensive devices.
To address this issue, we present a method for spatially upsampling spherical microphone arrays with a limited number of capsules. Our approach exploits a physics-informed neural network with Rowdy activation functions, leveraging physical constraints to provide high-order microphone array signals, starting from low-order devices.
Results show that, within its domain of application, our approach outperforms a state of the art method based on signal processing for spherical microphone arrays upsampling.
\end{abstract}
\vspace{-0.75mm}
\vspace{-1.1mm}
\section{Introduction}
\label{sec:introduction}
Sound field capturing and reproduction are key components of spatial audio \cite{cobos2022overview, rafaely2022spatial}, an ever-expanding field of research essential for applications such as virtual and augmented reality \cite{vorlander2020auralization, zhang2017surround} and teleconferencing \cite{miotello2024homula, alexandridis2013capturing}.
In this context, \acrfullpl{sma} are useful tools that feature multiple microphone capsules arranged around a sphere and are usually exploited for capturing the spatial characteristics of a sound field, which can then be rendered via headphones or loudspeaker arrays \cite{cobos2022overview, rafaely2022spatial}.
In particular, \acrshortpl{sma} allows for a efficient estimation of the spherical harmonics representation of the sound field \cite{rafaely2004analysis}, which is often exploited for different tasks including source localization \cite{cobos2023acoustic, hu2023generalized, olivieri2024acoustic}, separation \cite{mitsufuji2020multichannel, pezzoli2024spherical} and sound field separation \cite{pezzoli2022sparsity, fahim2017sound}. 

In practical scenarios, the number of capsules in a \acrshort{sma} imposes significant constraints on the spherical harmonics-domain representation of the captured sound field.
In fact, these devices sample sound fields at a limited number of points on the sphere's surface, potentially causing spatial aliasing and errors in encoding the acoustic field's spatial characteristics \cite{rafaely2004analysis, abhayapala2002theory, meyer2008handling}.
Therefore, it is necessary to limit the spherical harmonics expansion order to achieve an aliasing-free representation.
To overcome this challenge, researchers have explored spatial upsampling techniques for \acrshortpl{sma} \cite{lubeck2023spatial, xia2023upmix, bernschutz2012bandwidth}: finer spatial sampling shifts aliasing to higher frequencies, improving the spatial resolution of arrays with fewer sensors and offering a cost-effective solution without compromising performance.

In \cite{lubeck2023spatial}, the authors exploit a signal processing technique to perform upsampling by adding virtual microphones at locations between real microphones, achieving notable performances.
The virtual signals are generated by interpolating the measured signals, taking into account the differences in amplitude and time between neighboring signals.
Data-driven techniques, particularly those designed for spatial audio data processing, also show great promise in enhancing the spatial resolution of sound fields, effectively addressing tasks like sound field reconstruction or upsampling \cite{miotello2024reconstruction, karakonstantis2023generative, ronchini2024room}.
In \cite{xia2023upmix}, the authors propose a deep learning method, based on a generative adversarial network, for upsampling B-format room impulse responses, a task closely related to \acrshortpl{sma} upsampling.
However, this kind of techniques, relies its success on long training processes carried out on extensive datasets, which are often difficult to obtain.

Recently, many data-driven approaches, such as the deep-prior approach \cite{miotello2023deep, pezzoli2022deep, kong2020deep}, have attempted to address issues related to the limited amount of data available for training the models.
To this end, a notable recent approach involves integrating physics principles into neural network architectures, resulting in the development of physics-informed neural networks (PINNs) \cite{pezzoli2023implicit, olivieri2024physics, ma2023physics}.
The core idea is to ensure the network's output follows the partial differential equations (PDEs) governing the system under analysis.
In particular, in audio signal processing, these models leverage the wave equation's fundamental role in governing all sound fields.
This is achieved by exploiting the automatic differentiation framework, inherent in every neural network training process, for PDE computation.
PINNs proved to be effective when working with \acrshortpl{sma}, allowing authors to overcome the spherical Bessel function null with open arrays \cite{ma2023circumvent} and to estimate the sound field around a rigid sphere\cite{chen2023sound}.

In this paper, we propose a novel method for spatial upsampling of spherical microphone arrays, using physics-informed neural networks with Rowdy activation functions.
Our approach aims to achieve superior spatial resolution enhancement while maintaining computational efficiency and generalization capabilities.
Through empirical evaluation and comparative analysis, we demonstrate the effectiveness and robustness of our method in addressing the challenges associated with spatial upsampling of SMAs.
The paper is organized as follows. In Section~\ref{sec:problem}, we describe the SMA upsampling problem and the data model. In Section~\ref{sec:method}, we detail the proposed method. In Section~\ref{sec:results}, we present the results of our method, comparing them with those obtained using the technique from \cite{lubeck2023spatial}. Finally, in Section~\ref{sec:conclusion}, we draw our conclusions and suggest potential future expansions.

\section{Problem formulation}%
\label{sec:problem}%
\subsection{Data model}\label{sec:data_model}%
Let us consider a \acrshort{sma} of radius $R$, composed of $Q$ microphones located at spherical coordinates $\bm{r}_q = [R, \theta_q, \phi_q]^T \in \mathcal{Q}$, where $|\mathcal{Q}| = Q$.
The microphone signals acquired by the \acrshort{sma} can be encoded in the spherical harmonics domain as \cite{pezzoli2022sparsity}
\begin{equation}\label{eq:shd}
    C_{nm} = \frac{1}{b_n(kR)}
         \sum_{q \in \mathcal{Q}} p(\bm{r}_q,k) Y_{nm}^\ast(\theta_q, \phi_q),
\end{equation}
where $k = \frac{\omega}{c}$ is the wave number at angular frequency $\omega$ and speed of sound in air $c$, $p(\bm{r}_q, k)$ is the sound pressure at position $\bm{r}_q$. 
The spherical harmonic $Y_{nm}(\cdot)$ is defined as
\begin{equation}\label{eq:sh}
    Y_{nm}(\theta, \phi) = \sqrt{\frac{(2n+1)}{4\pi} \frac{(n-m)!}{(n+m)!}}
        P_{nm}(\cos{\theta}) e^{im\phi},
\end{equation}
where $n$ is referred to as the order of the spherical harmonic, and $m$ is referred to as its degree, $i = \sqrt{-1}$ is the imaginary unit and $P_{nm}(\cdot)$ is the associated Legendre polynomial of integer order $n$ and degree $m$.
The term $b_n(\cdot)$ in \eqref{eq:shd} is defined accordingly to the array enclosure type as \cite{williams1999fourier}
\begin{equation}
    b_n(kR) = 
    \begin{cases} 
      j_n(kR) & \text{if open sphere,} \\
      j_n(kR) - \frac{j'_n(kR)}{h'_n(kR)}h_n(kR) & \text{if rigid sphere,}
   \end{cases}
\end{equation}
where $j_n(\cdot)$ is the $n$th order spherical Bessel function of the first kind and $h_n(\cdot)$ is the $n$th order spherical Hankel function of the first kind.

In the scenario of an interior field problem, it is then possible to express the pressure value at an arbitrary point $\bm{r} = [r, \theta, \phi]^T$, by means of the spherical harmonics expansion \cite{abhayapala2002theory} as
\begin{equation}\label{eq:ishd}
    p(\mathbf{r}, k) = \sum_{n=0}^{\infty} \sum_{m=-n}^{n}
        C_{nm}(k) j_n(kr) Y_{nm}(\theta, \phi),
\end{equation}
where $C_{nm}$ are a set of harmonic coefficients, not depending on position $\bm{r}$, computed as in \eqref{eq:shd}.
Essentially, Equation \eqref{eq:shd} retrieves expansion coefficients $C_{nm}$ by sampling the sound field on a sphere of radius $R$, at point locations denoted by $\bm{r}_q$, which can then be used to retrieve pressure values at arbitrary points $\bm{r}$, using \eqref{eq:ishd}.

The spatial sampling of the sound field described in \eqref{eq:shd} can lead to aliasing, and the spacing between sampling points determines the critical frequency band where the contribution of aliasing artifacts prevail \cite{bernschutz2012bandwidth}.
Moreover, the summation in \eqref{eq:ishd}, can be in practice calculated only up to a certain order $N < \infty$, resulting in a truncation of the natural spherical harmonics expansion of the sound field \cite{lubeck2023spatial}.
Depending on the sampling scheme \cite{rafaely2004analysis}, a minimum of $Q=(N+1)^2$ sampling points is needed to resolve an expansion of order $N$ \cite{bernschutz2012bandwidth}, and generate an approximation $\Tilde{p}(\mathbf{r}, k) \approx p(\mathbf{r}, k)$ of the real sound field.
As a rule of thumb, a band limited expansion leads to a negligible aliasing if $kr \leq N$ \cite{bernschutz2012bandwidth, pezzoli2022sparsity}.
In order to limit aliasing while keeping the same radius $r$ it is possible to increase the number of sampled points $Q$ on the sphere \cite{meyer2008handling}.

\subsection{Spatial upsampling of \acrshortpl{sma}}
Let us consider the \acrshort{sma} presented in Section~\ref{sec:data_model}.
Given an arbitrary point on the surface of the sphere $\bm{r}_s = [R, \theta, \phi]^T$, the goal of \acrshort{sma} upsampling is to find a function that accurately retrieves the pressure field $\Tilde{p}(\bm{r}_s, k)$ based on a restricted set of observations $\Tilde{p}(\bm{r}_q, k)$.
In particular, the upsampling task can be interpreted in the framework of inverse problems, in which we aim at finding
\begin{equation}\label{eq:inverse_problem1}
\Tilde{p}(\bm{r}_s, k) = f_{\bm{\theta}}(\bm{r}_s) \approx p(\bm{r}_s, k),
\end{equation}
which is an estimate of $p(\bm{r}_s, k)$, computed using function $f_{\bm{\theta}}(\cdot)$ that estimates pressure values at locations $\bm{r}_s$ using parameters $\bm{\theta}$.
The solution to the upsampling problem can be retrieved by an optimization process
\begin{equation}\label{eq:inverse_problem}
    \bm{\theta}^* = \underset{{\bm{\theta}}}{\text{argmin}}\,\,J\left(\bm{\theta}\right) = E \left( f_{\bm{\theta}}(\bm{r}_q), \Tilde{p}(\bm{r}_q, k) \right),
\end{equation}
where $E(\cdot)$ is a data-fidelity term, e.g., the mean squared error (MSE), between the estimated and available data.
It is worth noting that in \eqref{eq:inverse_problem}, the evaluation of the reconstruction error is performed only on the observed locations $\bm{r}_q$. 
However, in order to takle the upsampling task, $f$ must be able to provide a meaningful estimate also in locations on the \acrshort{sma} surface, different from the available ones, i.e., $\bm{r}_s$. 
Therefore, the solution to the optimization problem \eqref{eq:inverse_problem} must be constrained using regularization strategies on the upsampled pressure field $\Tilde{p}(\bm{r}_s, k)$. 
In the context of audio signal processing, typical regularization techniques include compressed sensing frameworks based on assumptions about the signal model \cite{damiano2024compressive}, such as plane and spherical wave expansions \cite{koyama2019sparse}, as well as deep learning approaches \cite{lluis2020sound, karakonstantis2023generative, miotello2024reconstruction}.

\begin{figure*}[ht]
\centering
\begin{minipage}[t]{0.48\textwidth}
  \centering
  \centerline{\begin{tikzpicture}%
\begin{axis}[%
    xlabel={Time},%
    ylabel={Amplitude},%
    x unit = \second,
        ymin=-0.08,
    ymax=0.17,
    axis x line=bottom,%
    axis y line=left, %
    grid, %
    height=4cm, %
    width=\columnwidth, %
    enlarge x limits=0.08,%
    enlarge y limits=0.08,%
    style={font=\normalsize},%
    legend columns=2,%
    legend style={at={(0.65,0.78)},anchor=south,font=\small},%
    log ticks with fixed point,
    legend cell align={left}
]

\addplot[line width=0.2mm, blue] table {results/rir_gt.txt};%
\addlegendentry{Ground truth}%

\addplot[densely dotted, line width=0.4mm, orange] table {results/rir_sarita.txt};%
\addlegendentry{SARITA} %

\end{axis}%
\end{tikzpicture}}
  \centerline{(a)}
  \vspace{.5em}
\end{minipage}
\hfill
\begin{minipage}[t]{0.48\textwidth}
  \centering
\centerline{\begin{tikzpicture}%
\begin{axis}[%
    xlabel={Time},%
    ylabel={Amplitude},%
    x unit = \second,
    ymin=-0.08,
    ymax=0.17,
    axis x line=bottom,%
    axis y line=left, %
    grid, %
    height=4cm, %
    width=\columnwidth, %
    enlarge x limits=0.08,%
    enlarge y limits=0.08,%
    style={font=\normalsize},%
    legend columns=2,%
    legend style={at={(0.65,0.78)},anchor=south,font=\small},%
    log ticks with fixed point,
    legend cell align={left}
]

\addplot[line width=0.2mm, blue] table {results/rir_gt.txt};%
\addlegendentry{Ground truth}%

\addplot[densely dotted, line width=0.4mm, orange] table {results/rir_proposed.txt};%
\addlegendentry{Proposed} %

\end{axis}%
\end{tikzpicture}}
  \centerline{(b)}
  \vspace{.5em}
\end{minipage}
\caption{Comparison between ground truth RIR and RIRs reconstructed using (a) SARITA and (b) the proposed method, starting from 9 available \acrshort{sma} channels.}
\label{fig:rirs}
\end{figure*}
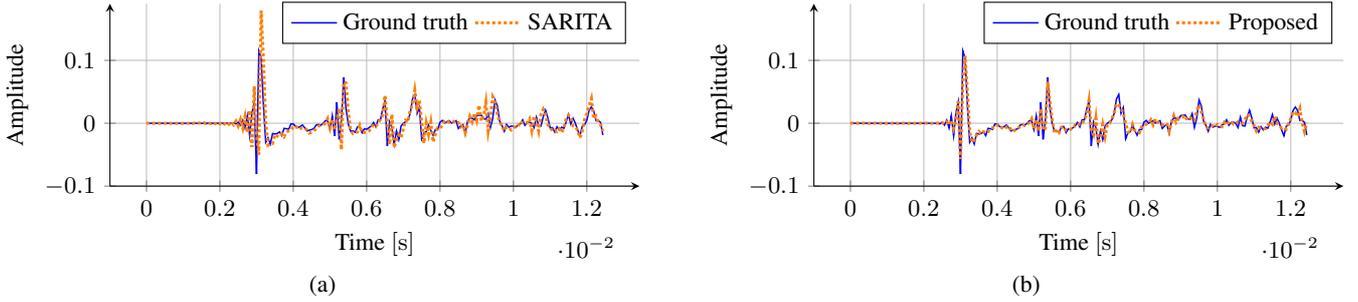%
\section{Method}
\label{sec:method}
In this work, we aim at solving the \acrshort{sma} upsampling problem in \eqref{eq:inverse_problem}, in order to retrieve $\Tilde{p}(\bm{r}_s, k) = f_{\bm{\theta}^\ast}(\bm{r}_s)$,
where function $f_{\bm{\theta}^\ast}(\cdot)$ represents a neural network with optimized learnable parameters $\bm{\theta}^\ast$ retrieved as in \eqref{eq:inverse_problem}.
In particular, following an approach similar to the one proposed in \cite{pezzoli2023implicit}, we adopt the architecture of a SIREN \cite{sitzmann2020implicit} neural network.
This model proved its efficacy in retrieving the so-called neural implicit representation of various kinds of signals, including audio. Initially designed as a \acrshort{mlp} with sinusoidal activation functions, in this work we enhance the SIREN model using Rowdy activation functions \cite{jagtap2022deep} which showed superior performance in \cite{ribeiro2023sound}.
In particular, the output to the $i$th layer of the neural network can be expressed as
\begin{equation}\label{eq:layer}
    \Lambda_i = \sigma_i(\mathbf{x}_i^T \bm{\theta}_i + \mathbf{b}_i),
\end{equation}
where $\mathbf{x}_i$ is the input coming from the previous layer, $\bm{\theta}_i$ and $\mathbf{b}_i$ are respectively the weights and biases relative to the $i$th layer, and $\sigma_i$ is the Rowdy activation function.
More precisely, $\sigma_i$ can be expressed as
\begin{equation}\label{eq:activation}
    \sigma_i(x) = \sin(\omega_0 x) + \sum_{w = 1}^W n_w\sin(\alpha_w x),
\end{equation}
where $\omega_0$ is an initialization hyper-parameter of the SIREN, $n_w$ is a scaling factor and $\alpha_w$ is a multiplicative factor changing the frequency of the $\sin$ function.
Both factors are optimized during training and are initialized as $n_w = 1$ and $\alpha_w = w$.
Rowdy activation functions were introduced in \cite{jagtap2022deep} to capture the high-frequency components within the target objective function.
They achieve this by integrating high-frequency sinusoidal fluctuations into the underlying base activation function.
It has been demonstrated in \cite{jagtap2022deep} that they facilitate quicker learning and yield improved outcomes, particularly in the context of \acrshortpl{pinn}.

The adopted SIREN architecture can thus be described as a composition of $L$ layers as
\begin{equation}
    f_{\bm{\theta}}(\mathbf{x}) = (\Lambda_L \circ \Lambda_{L-1} \circ \dots \circ \Lambda_1)(\mathbf{x}),
\end{equation}
where $\mathbf{x}$ is the input to the network and $\bm{\theta}$ the set of learnable parameters.
Following to the neural implicit representation paradigm, the SIREN model accepts the signal domain as input, specifically the positions $\bm{r}_s$, and outputs an estimate of signals $\Tilde{p}(\mathbf{r}_s, k)$ \eqref{eq:inverse_problem1}, across the entire considered frequency band.
Thus, the network's role is to produce a parameterized description of the signals through the parameters of the \acrshort{mlp}.

Similarly to \cite{pezzoli2023implicit, karakonstantis2023room, olivieri2024physics}, in this study, we adopt a \acrshort{pinn} approach for training the SIREN model.
In fact, by exploiting the observations only as training target, there is no guarantee that the solution adheres to the physical law governing the underlying problem, i.e., the Helmholtz equation.
Therefore, the output of \acrshortpl{pinn} is constrained to fit the solutions of PDEs representing a physical model of the acoustic system, thus improving the results.
Such a physically meaningful regularization is included in the loss function defined in a similar way to \cite{song2023simulating} as
\begin{equation}\label{eq:loss}
\begin{aligned}
    \mathcal{L} &= \frac{1}{Q} \sum_{\mathbf{r}_q \in \mathcal{Q}} \| \Hat{p}(\mathbf{r}_q, k) - \Tilde{p}(\mathbf{r}_q, k) \|_2^2 +\\
    &+ \lambda \frac{1}{S} \sum_{s=1}^S
    \| [\nabla^2 \Hat{p}_\Re(\mathbf{r}_s, k) + i\nabla^2 \Hat{p}_\Im(\mathbf{r}_s, k)] +\\
    &+ k^2 \Hat{p}(\mathbf{r}_s, k) \|_2^2,
\end{aligned}
\end{equation}
where $\|\cdot\|_2^2$ is the $\ell_2$-norm, $\Hat{p}$ and $\Tilde{p}$ denote respectively the network estimate and the measured sound pressure, $\Hat{p}_\Re$ and $\Hat{p}_\Im$ represent, respectively, the real and imaginary parts of $\hat{p}$, while $S$ is the number of points $\bm{r}_s$ at which we are evaluating the Helmholtz equation.
The first term of \eqref{eq:loss} represents the distance between the predicted and the available signals (i.e., the MSE) and ensures that the network output matches the observations.
The second term, instead, corresponds to the PDE loss given by the Helmholtz equation weighted by parameter $\lambda$.
Integrating the PDE loss leads to a regularized solution, as the output aligns with the fundamental physical equation.
Post-training, the model facilitates the retrieval of microphone signals at both available and absent positions of the \acrshort{sma} by simply inputting the locations $\mathbf{r}_s$ into the network.
\section{Evaluation}
\label{sec:results}
\subsection{Experiments setup}
We evaluate the performance of our method for the \acrshort{sma} upsampling task of RIR signals acquired using the mh acoustics' Eigenmike EM32 \cite{schneiderwind2019data} and considering a sampling frequency $\text{\textit{fs}} = \SI{16}{k\hertz}$.
The environment in which the RIR is recorded is a small conference room with dimensions $\SI{10.3}{\meter} \times \SI{5.8}{\meter} \times \SI{3.1}{\meter}$ and reverberation time $\operatorname{T_{60}} = \SI{0.63}{\second}$.

Similarily to \cite{ma2023physics}, our model is composed of two parallel SIREN architectures with Rowdy activation functions.
The first one accounts for the real part of the input, the other for the imaginary part, and are jointly trained sharing the same loss function \eqref{eq:loss}.
The outputs of the two networks are then combined together to generate the upsampled complex sound field.
Both networks are composed of L = 4 hidden layers of $512$ neurons each, other than an input and one output layers.
The initialization frequency $\omega_0$ in \eqref{eq:layer} is set to $1$ for the first layer, while $\omega_0 = 5$ for the hidden layers, and parameter $W$ in \eqref{eq:activation} set to $6$.
The network is trained for $10000$ iterations using the Adam optimizer, with learning rate initially set to \num{1e-4} and gradually lowered using cosine annealing to prevent overfitting.
The PDE weight parameter in \eqref{eq:loss} has ben empirically set to $\lambda = \num{1e-12}$.

We compare our method with SARITA \cite{lubeck2023spatial}, a technique which performs \acrshort{sma} upsampling by leveraging variations in amplitude and time among neighboring signals, and computes the interpolation accordingly.
We evaluate the performance of the two techniques in terms of normalized mean squared error (NMSE) defined in time domain as
\begin{equation}\label{eq:nmse}
    \operatorname{NMSE} =
    10\log_{10} \frac{1}{S}
    \sum_{s=1}^S \frac{\|
    \Hat{\mathbf{p}}(\mathbf{r}_s, t) - \Tilde{\mathbf{p}}(\mathbf{r}_s, t)
    \|^2_2}
    {\|\Tilde{\mathbf{p}}(\mathbf{r}_s, t)
    \|^2_2},
\end{equation}
where $\Hat{\mathbf{p}}$ and $\Tilde{\mathbf{p}}$ are respectively the estimated and the measured sound field in the time domain.
Evaluation points $\bm{r}_s$ correspond to the positions of the microphone capsules on the mh acoustics’ Eigenmike EM32.
For both methods, upsampling is performed considering of having access to $Q = \{4, 9, 16, 25\}$ equidistant microphones out of the total $32$ present in the given \acrshort{sma}.
These numbers correspond to the minimum points required to represent a $1^\text{st}$, $2^\text{nd}$, $3^\text{rd}$, and $4^\text{th}$ spherical harmonics order sound field, respectively.

\subsection{Results and discussion}

Results are displayed in Table~\ref{tab:res}.
As a first experiment to initially validate the adopted solution, we compared results obtained using three different architectures.
In particular, the first one (referred to as SIREN in Table~\ref{tab:res}) is the plain SIREN \cite{sitzmann2020implicit} architecture, with sinusoidal activation functions and solely trained using the data loss term in \eqref{eq:loss}. The second one (referred to as SIREN + PDE in Table~\ref{tab:res}) is the very same SIREN architecture trained considering the entire loss function in \eqref{eq:loss}.
Finally, the third one (referred to as Proposed in Table~\ref{tab:res}) is the architecture we are presenting in this work, characterized by Rowdy activation functions \cite{jagtap2022deep} and the loss function described in \eqref{eq:loss}.
Across all considered numbers of available channels $Q$, the proposed solution consistently outperforms the other models, with a minimum performance improvement of \SI{0.88}{\decibel} when considering $Q=4$ channels, and a maximum improvement of \SI{2.80}{\decibel} when considering $Q=9$ channels.
\begin{table}[h]
\centering
\caption{Mean NMSE with respect to the number of available channels in the \acrshort{sma}.}
\begin{tabular}{l||c|c|c|c}
\multicolumn{1}{l}{}     & \multicolumn{4}{c}{Mean NMSE}                                                                                                                                                               \\ 
\hhline{=:t:====}
$Q$ & 4                                            & 9                                            & 16                                           & 25                                             \\ 
\hhline{=::====}
SARITA                    & \textcolor[rgb]{0.2,0.2,0.2}{-0.65}          & \textcolor[rgb]{0.2,0.2,0.2}{-2.6}          & \textcolor[rgb]{0.2,0.2,0.2}{-4.9}          & \textcolor[rgb]{0.2,0.2,0.2}{-5.57}           \\ 
\hline
SIREN                    & \textcolor[rgb]{0.2,0.2,0.2}{-1.17}          & \textcolor[rgb]{0.2,0.2,0.2}{-2.60}          & \textcolor[rgb]{0.2,0.2,0.2}{-5.76}          & \textcolor[rgb]{0.2,0.2,0.2}{-10.92}           \\ 
\hline
SIREN + PDE              & \textcolor[rgb]{0.2,0.2,0.2}{-1.71}          & \textcolor[rgb]{0.2,0.2,0.2}{-4.97}          & \textcolor[rgb]{0.2,0.2,0.2}{-6.38}          & \textcolor[rgb]{0.2,0.2,0.2}{-11.13}           \\ 
\hline
Proposed                 & \textcolor[rgb]{0.2,0.2,0.2}{\textbf{-2.05}} & \textcolor[rgb]{0.2,0.2,0.2}{\textbf{-5.40}} & \textcolor[rgb]{0.2,0.2,0.2}{\textbf{-6.83}} & \textcolor[rgb]{0.2,0.2,0.2}{\textbf{-12.44}} 
\end{tabular}
\label{tab:res}
\end{table}

%

The proposed solution also consistently outperforms SARITA \cite{lubeck2023spatial} across all considered configurations, achieving a minimum improvement of \SI{1.40}{\decibel} with $Q=4$ channels and a maximum improvement of \SI{6.87}{\decibel} with $Q=25$ channels out of 32. 
These findings suggest that leveraging knowledge of the underlying physics enables our method to produce more accurate results.
This can be noticed when examining Fig.~\ref{fig:rirs}, where RIRs reconstructed using both SARITA and the proposed method, with access to $9$ available SMA channels, are compared with the ground truth RIR. 
Our solution appears to reconstruct the original RIR , closely following the peaks without overestimating the amplitude values.

It is worth noting that signal processing methods such as SARITA do not require a training phase, enabling faster upsampling.
Conversely, methods exploiting PINNs require an initial training phase tailored to the specific setup under consideration. 
However, once this training is completed, the estimation of the pressure value at a new position $\bm{r}_s$ becomes immediate.



\section{Conclusion}
\label{sec:conclusion}

In this paper we have proposed a PINNs-based approach for the spatial upsampling of \acrshortpl{sma}.
Specifically, we considered SIREN, a MLP with sine activation functions, already successfully exploited in many audio related tasks.
We enhanced the SIREN model by incorporating Rowdy activation functions, which help the network in capturing high-frequency components within the target objective function and facilitate the convergence towards meaningful solutions.
Through an experimental campaign on measured data, we validate the adopted solution and compare its performances with a signal processing-based \acrshort{sma} upsampling method, demonstrating the effectiveness of
the our approach.
The obtained results encourage us to further explore the topic, by incorporating additional physical constraints more closely related to our application domain and testing the proposed method on more challenging scenarios.
\bibliographystyle{ieeetr}
\bibliography{refs}

\end{document}